\documentclass[twocolumn,aps,prb,superscriptaddress]{revtex4}

\usepackage{graphicx,amsfonts,amssymb,amsmath,bm,empheq,color,hyperref}

\begin{document}

\title{Composite topological structure of domain walls in synthetic antiferromagnets}

\author{A. G. Kolesnikov}
\affiliation{School of Natural Sciences, Far Eastern Federal University, Vladivostok 690950, Russia}
\author{V. S. Plotnikov}
\affiliation{School of Natural Sciences, Far Eastern Federal University, Vladivostok 690950, Russia}
\author{E. V. Pustovalov}
\affiliation{School of Natural Sciences, Far Eastern Federal University, Vladivostok 690950, Russia}
\author{A. S. Samardak}
\affiliation{School of Natural Sciences, Far Eastern Federal University, Vladivostok 690950, Russia}
\affiliation{Center for Spin-Orbitronic Materials, Korea University, Seoul, 02841, Republic of Korea}
\affiliation{National Research South Ural State University, Chelyabinsk, 454080, Russia}
\author{L. A. Chebotkevich}
\affiliation{School of Natural Sciences, Far Eastern Federal University, Vladivostok 690950, Russia}
\author{A. V. Ognev}
\affiliation{School of Natural Sciences, Far Eastern Federal University, Vladivostok 690950, Russia}
\author{Oleg A. Tretiakov}
\email{olegt@imr.tohoku.ac.jp}
\affiliation{Institute for Materials Research, Tohoku University, Sendai 980-8577, Japan}
\affiliation{School of Natural Sciences, Far Eastern Federal University, Vladivostok 690950, Russia}

\date{\today}

\begin{abstract}
We experimentally study the structure and dynamics of magnetic domains in synthetic antiferromagnets based on Co/Ru/Co films.  Dramatic effects arise from the interaction among the topological defects comprising the dual domain walls in these structures. Under applied magnetic fields, the dual domain walls propagate following the dynamics of bi-meronic (bi-vortex/bi-antivortex) topological defects built in the walls.  Application of an external field triggers a rich dynamical response: The propagation depends on mutual orientation and chirality of bi-vortices and bi-antivortices in the domain walls. For certain configurations, we observe sudden jumps of composite domain walls in increasing field, which are associated with the decay of composite skyrmions.  These features allow for enhanced control of domain-wall motion in synthetic antiferromagnets with the potential of employing them as information carriers in future logic and storage devices.

\end{abstract}

\maketitle

In recent years there was an enormous interest in complex magnetic nanostructures due to a wide variety of novel effects arising from their magnetic properties and the corresponding prospects for their applications for future spintronic memory, logic, and sensors.\cite{Parkin:racetrack, Allwood01, Allwood02} In particular, significant efforts have been focused on studying statics and dynamics of domain walls (DWs) in ferromagnetic wires \cite{Nakatani03, Yamaguchi04, Tatara04, Zhang04, Beach05, KlauiPRL05, Thomas2006, Duine07, Tserkovnyak2008, Moriya08, Tretiakov2008, BeachPRL09, IlgazPRL10, Krivorotov10, Tretiakov_DMI, Thomas2010, Tretiakov2012} and thin films with perpendicular magnetic anisotropy. \cite{Thiaville2012, Beach2013, Parkin2013,  Brataas2013, Ohno2014}  In ferromagnet/heavy-metal multilayer systems, very high velocities have been observed for chiral DWs.\cite{Miron2011, Emori2013, Ono2016} Topology of the magnetic textures has been shown to play an essential role for the potential applications, leading to new ideas of employing compact topological spin textures -- skyrmions -- for the racetrack memory prototypes. \cite{Fert2013} This boosted various observations of skyrmions in multilayer materials, e.g. using a constriction in a trilayer system to create skyrmions at room temperatures \cite{Hoffmann2015} and demonstrate their current-driven dynamics. \cite{Hoffmann2016_Sk} The observation of room-temperature magnetic skyrmions has been also established in ultrathin metallic ferromagnetic (FM) multilayers, such as Pt/Co/Ta and Pt/CoFeB/MgO, \cite{Beach2016_Sk} and the skyrmion Hall effect has been revealed by time-resolved X-ray microscopy. \cite{LitziusNPhys2017}

More recently magnetic bilayer-skyrmions experiencing no skyrmion Hall effect have been predicted in antiferromagnetically coupled FM bilayers,\cite{Ezawa2016_biSk} similarly to real antiferromagnetic (AFM) systems \cite{Barker2016}. The absence of skyrmion Hall effect allows for better skyrmion motion control in future nanodevices, and antiferromagnetically coupled FM bilayer systems -- synthetic antiferromagnets  (SAFs) \cite{Grunberg1986, Parkin1991} -- offer a particular advantage, since this effect is easier to observe in them and employ for spintronic applications. SAFs were already demonstrated to have serious advantages for applications: high speeds of the DWs have been observed in them\cite{Parkin2015} and the magnetic microscopy has shown the topological stability of homochiral N{\'e}el DWs. \cite{Marrows_SAF2015} As we show below, SAFs possess much richer diversity of magnetic topological defects, especially for materials with in-plane anisotropy, however this fact began to be explored experimentally only very recently. 
Earlier most of theoretical and experimental studies have been devoted to the films with the indirect ferromagnetic coupling, \cite{Slonczewski1965}  where e.g. asymmetric domain nucleation and unusual magnetization reversal has been observed in ultra-thin Co films with perpendicular anisotropy. \cite{Iunin07} 

\begin{figure}
\includegraphics[width=0.95\linewidth]{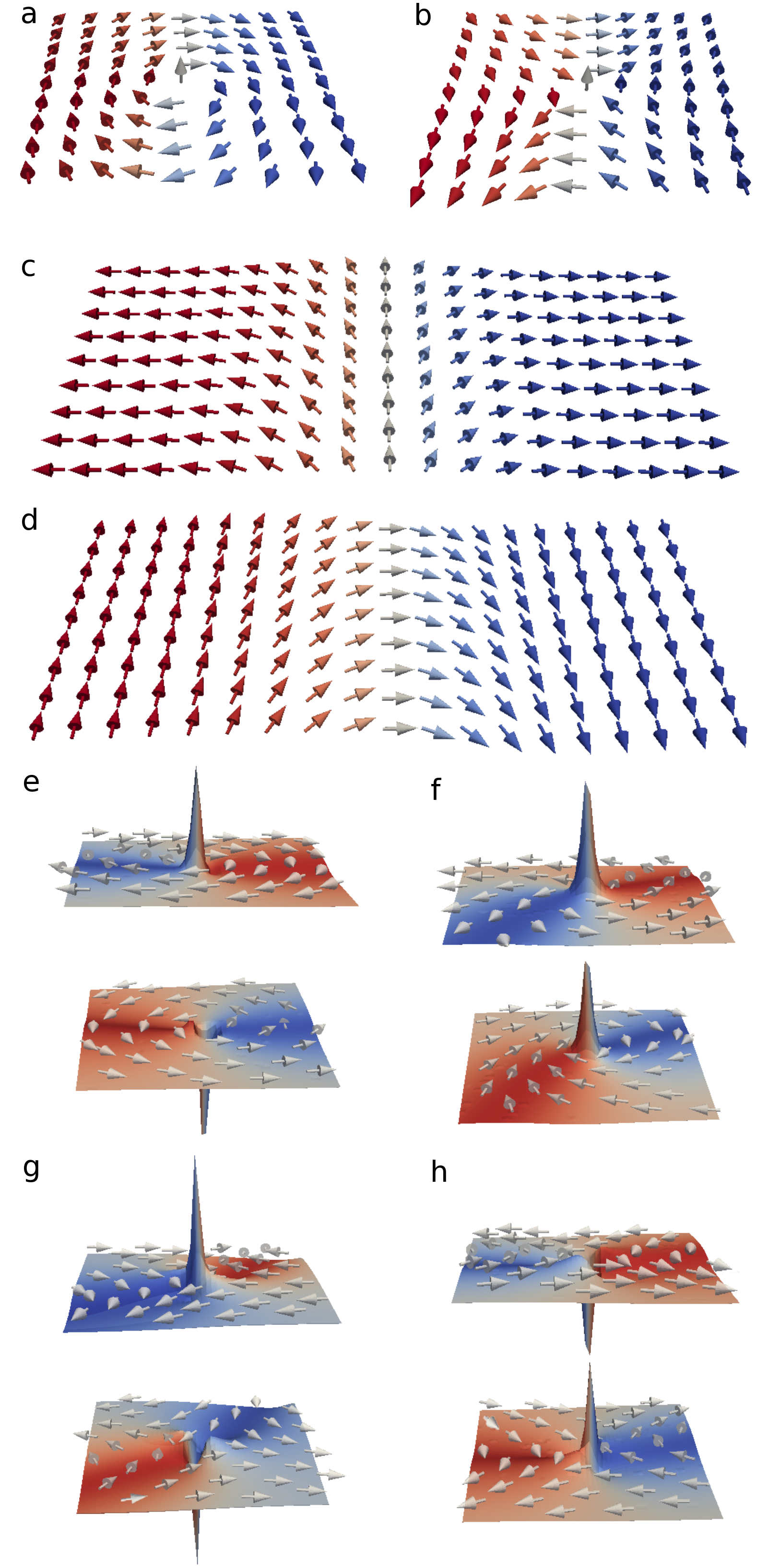}
\caption{Domain wall building blocks: (a) Vortex, (b) Antivortex,  (c) Transverse domain wall, (d) N{\'e}el domain wall. (e) -- (h) Vortex (antivortex) pairs with polarities $p = + 1$ ($-1$) in the top and bottom layers.
(e) Vortex pair with $p=+1$, $p=-1$; (f) Vortex pair with $p=+1$, $p=+1$; (g) Antivortex pair with $p=+1$, $p=-1$; and (h) Antivortex pair with $p=-1$, $p=+1$.}
\label{fig:DW_blocks}
\end{figure}

In this article we reveal the composite structure of magnetic domains in SAFs with in-plane anisotropy based on Co/Ru/Co films. We find that in this system both dual N{\'e}el (NDW) and transverse (TDW) domain walls are present and connected by dual topological defects -- bi-merons, which are pairs of coupled vortices or antivortices in the upper and lower layers of SAF, see Fig.~\ref{fig:DW_blocks}(e) -- (h). We show that under the influence of magnetic field the magnetization reversal processes in the SAF domains are determined by the dynamics of these composite DWs coupled in the two layers. The controllable transformations of the DW types we observe in the process of magnetization reversal are demonstrated to be governed by the motion and annihilation of magnetic bi-merons. Our experimental results are supported by both micromagnetic simulations and general topological arguments.

\section*{Results}

\textbf{Magnetization measurements.}  To investigate the magnetic properties of  Co/Ru/Co films with AFM exchange coupling we have measured the hysteresis loops by longitudinal magneto-optical Kerr effect with fields applied parallel to the plane of the film. The shape of the hysteresis loop at the first AFM maximum ($t_{\rm{Ru}} = 0.9$ nm) shows that the magnetization reversal is mainly dominated by the processes of magnetization rotation (see Fig.~1 of the Supplementary Material). In small fields, the hysteresis is negligible and coercivity $H_{\rm{c}} = 60$ Oe. In the field range from $\pm 0.3$ up to $\pm 0.9$ kOe the loop is disclosed. The value of the field, at which there is maximum disclosure of the hysteresis loop, is denoted by $H_{\rm{c2}}$. The measurements give $H_{\rm{c2}} = 0.735 \pm 0.015$ kOe, while the saturation field is $H_{\rm{s}} = 1.8$ kOe. 

The indirect exchange coupling constant between the Co layers coupled antiferromagnetically has been evaluated according to the expression: \cite{Demokritov1998}
\begin{equation}
J_{in} = -\frac{1}{2} H_s t_{\rm{Co}} M_s,
\end{equation}
where $M_s$ is the saturation magnetization.
Then for the film Co/Ru/Co we find $H_s = 2.2$ kOe, $M_s = 1260$ G, $t_{\rm{Co}} = 10$ nm and $J_{in}= -1.4$ erg/cm$^2$. The value of $M_s$ was taken accounting for the polycrystalline structure of the films.

\begin{figure*}[ht]
\includegraphics[width=0.99\linewidth]{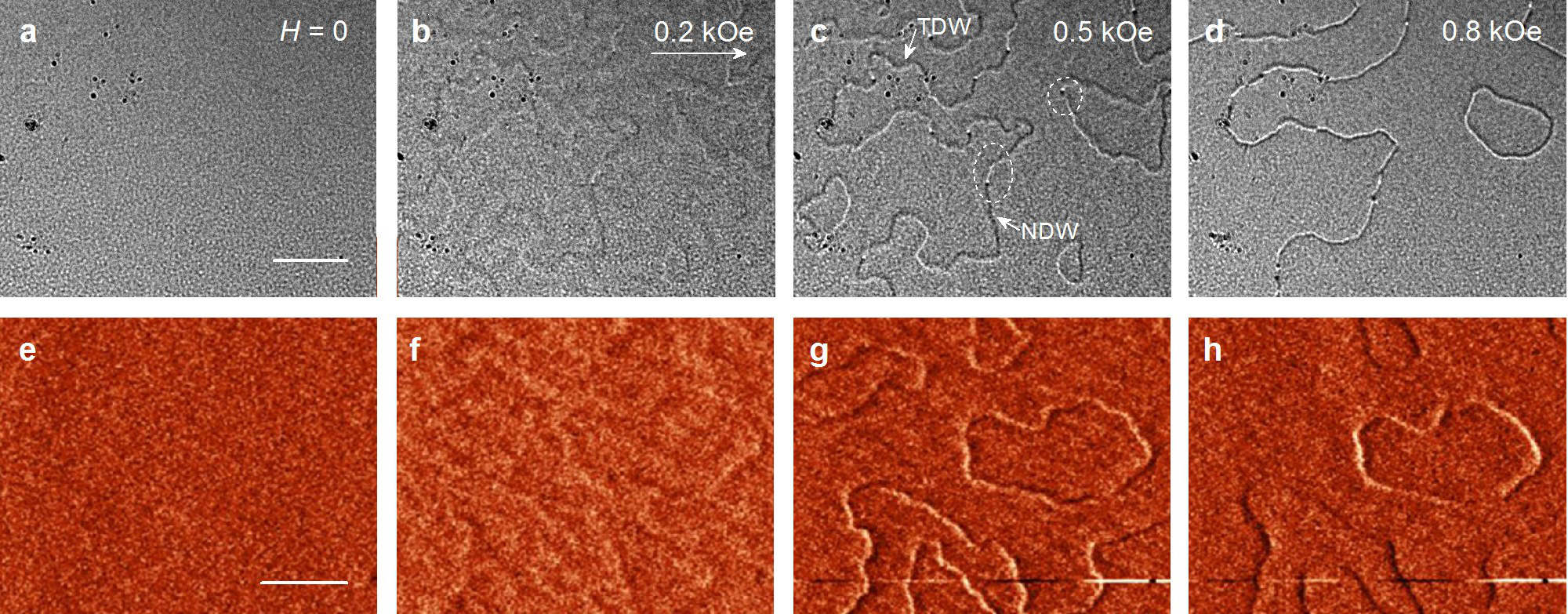}
\caption{(a) -- (d) The Lorentz microscopy images of the domain structure for the films Co/Ru(0.9 nm)/Co are shown for the fields from 0 to 0.8 kOe. (e) -- (h) The MFM images of the domain structure of a different part of the same film.}
\label{fig:2}
\end{figure*}

\textbf{Magnetic imaging.} Here we present the results for the Lorentz TEM and MFM studies of the magnetic film structure. Figure~\ref{fig:2} shows the image of the magnetic film structure of Co/Ru(0.9 nm)/Co in the magnetic fields from 0 to 0.8 kOe. Initially, the magnetic field of $H = 10$ kOe was applied perpendicular to the film (along $z$ axis). At $H = 0$ the contrast of the magnetic structure is absent. As magnetic field increases, the contours of domain boundaries start to appear on the  Lorentz TEM images (see Fig.~\ref{fig:2}(b), $H = 0.2$ kOe). A weak contrast of twisted magnetization ripples becomes visible. At $H = 0.5$ kOe, Fig.~\ref{fig:2}(c), the domain boundaries are already clearly visible. They wind and form closed domains of different sizes. This indicates that the indirect exchange coupling suppresses the induced magnetic anisotropy.

Let us next consider the specific features of the domain walls. For Co layers of 10 nm thickness we expect N{\'e}el type domain walls, with the magnetization reversal in the film plane. Figure~\ref{fig:2}(c) shows that the domain walls form two types of contrast: two neighboring walls (bright and dark lines very close together) and a single wall (bright or dark lines).
The domain walls of different contrast are separated by light and dark dots, as well as the areas without any contrast, they are marked by the dashed circles in Fig.~\ref{fig:2}(c). In field $H = 0.95$ kOe small closed domain collapses, most of the walls form bright and dark contrast, the areas with inverted contrast are almost gone.
In $H = 1.1$ kOe the contrast from the domain boundaries disappears (see Fig.~2 in the Supplementary Material), however the saturation in this case is not yet reached, $M_r/M_s \approx 0.65$ (see Fig.~1 in the Supplementary Material).

We conducted an additional study of the domain structure by MFM. The results are shown in Fig.~\ref{fig:2}(e) -- (h). In zero magnetic field, Fig.~\ref{fig:2}(e), the contrast of the magnetic structure is completely absent. With the increase of the field to $H = 0.2$ kOe the low contrast becomes visible due to the magnetization ripples and precursors of the domain walls. At $H = 0.3$ kOe the contrast of the domain walls appears. The domain walls form closed magnetization domains.

With increasing magnetic field up to $H = 0.8$ kOe position of the DWs is changed slightly. Figures~\ref{fig:2}(g) and (h) show that the contrast of the DWs changes from a bipolar (light and dark) to unipolar (dark or light). This indicates a change in the type of domain walls, e.g. from NDW to TDW. Moreover, it shows the presence of both NDWs and TDWs on these images (Fig.~3 in the Supplementary Material provides more MFM images in the range from -0.8 to 0.8 kOe).

\begin{figure}[ht]
\includegraphics[width=0.99\linewidth]{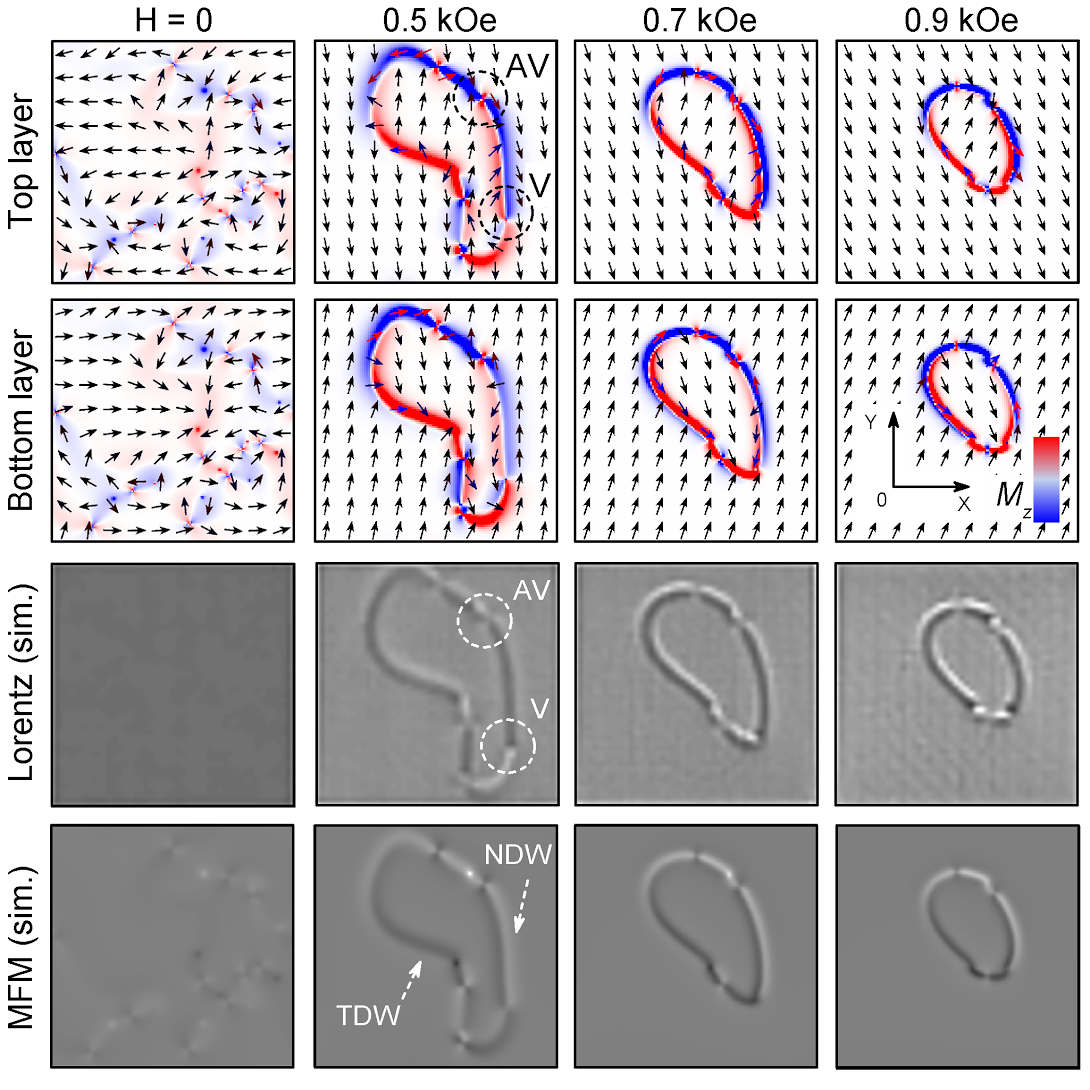}
\caption{Simulations: The domain structure of Co/Ru(0.9 nm)/Co film. From top to bottom: top-layer and bottom-layer magnetization structures, simulated using MALTS Lorentz microscopy image \cite{Walton2013} for the corresponding Co/Ru(0.9 nm)/Co magnetization structure, simulated MFM image in MuMax3.}
\label{fig:3}
\end{figure}

\textbf{Topological domain walls.} For the proper decoding of the domain structure and the type of domain boundaries we perform micromagnetic simulations using the software package MuMax3. \cite{MuMax} 
The simulation results are presented in Fig.~\ref{fig:3}. In demagnetized state at $H = 0$  a disordered array of bi-vortices and bi-antivortices is formed in the sample. The magnetization in the adjacent layers is antiparallel to each other. In spite of the fact that the film has induced magnetic anisotropy, the AFM interlayer coupling strongly diminishes its influence so that the magnetization distribution in AFM coupled bilayers becomes isotropic. The configuration of DWs shows the existence of this isotropy. 

With increasing magnetic field starting from $H = 0.3$ kOe the domain walls are already formed in the film. For field $H = 0.5$ kOe in Fig.~\ref{fig:3}  it is represented by a closed domain formed by different types of DWs. In the process of domain formation some of the bi-vortices and bi-antivortices annihilate, and the remaining ones belong to the formed DWs. 

One can see that the magnetization in the domains is at a small angle relative to the $y$ axis. The DWs, which are parallel to the magnetization, are the NDWs. The polarity of the NDWs in adjacent layers is different. The domain walls oriented perpendicular to the magnetization are the TDWs. The magnetization in the TDWs in adjacent layers is antiparallel to each other.
Because on the MFM simulated images can be seen that N{\'e}el DWs correspond to a bipolar contrast (light -- dark), a transverse DWs generate only unipolar (dark or light) contrast. In this case, TDWs appear as dark bands. From the Lorentz imaging it is difficult to make an unambiguous conclusion about the DW type. On the  Lorentz microscopy images in the field of 0.5 kOe, a NDW corresponds to unipolar contrast and TWD to a bipolar contrast, both of them are indicated in Fig.~\ref{fig:2}(c).

The DWs of different contrast are connected by a bi-vortex (V) or bi-antivortex (AV). \cite{Hubert98book} Figure~\ref{fig:DW_blocks} shows the structure of such bi-vortices and bi-antivortices, which connect NDW and TDW in the film. The analysis of our micromagnetic simulations shows that the bi-vortices can have the same or opposite polarities, $V_{\rm{P}}$ or $V_{\rm{AP}}$, respectively. Meanwhile, the bi-antivortices always have the opposite polarities in the upper and lower layers. The orientation of the magnetic moments in the adjacent layers of both bi-vortex and bi-antivortex is antiparallel.

It follows from the simulated images of Lorentz microscopy (see Fig.~\ref{fig:3}) that the vortices of different chirality can be seen as the adjacent bright and dark areas in the domain wall, whereas antivortices show no contrast. They are shown by the dashed circles in Fig.~\ref{fig:3}.
On the simulated MFM images, the vortex is seen as an area with alternating contrast from dark to light, and the antivortex is seen as a narrow area of the same contrast (dark or light) passing across the wall. Antivortices are always formed between N{\'e}el  and transverse DWs.

\begin{figure}[ht]
\includegraphics[width=0.99\linewidth]{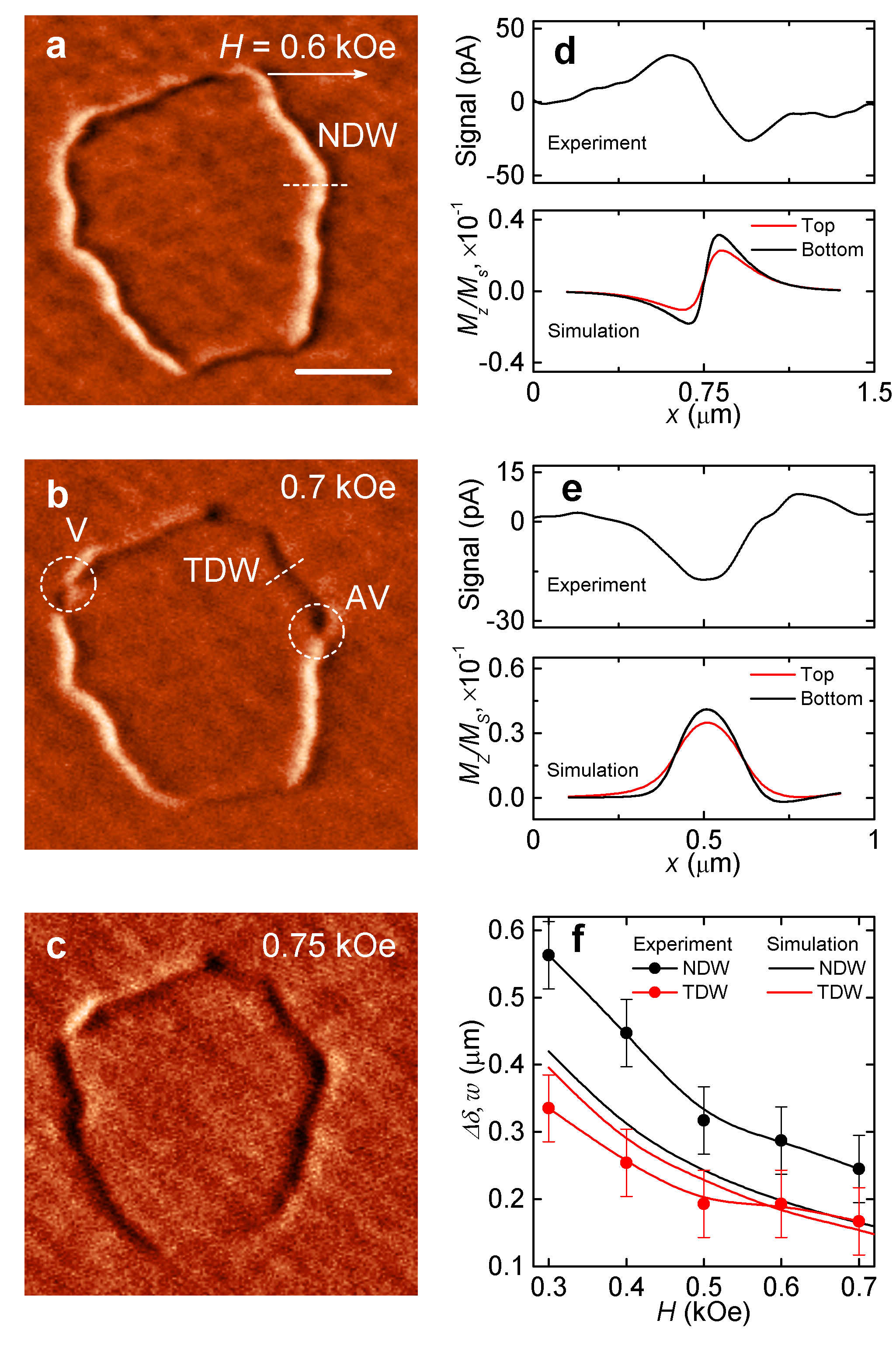}
\caption{MFM images of domain structure of the Co/Ru(0.9 nm)/Co film, obtained in fields directed in the plane of, (a) - (c) corresponds to the fields $H = 0.6$, 0.7, and 0.75 kOe; experimental and simulated profiles of N{\'e}el domain wall (d) and transverse DW (e). (f) The simulated DW width ($w$) and experimental DW profile width ($\Delta\delta$) as a function of external magnetic field.}
\label{fig:4}
\end{figure}

Figure~\ref{fig:4} shows the high-resolution images of MFM closed domain. The scanning is performed in a magnetic field pointing along $x$ axis. One can see that in field $H = 0.6$ kOe the domain walls are of two types: N{\'e}el domain walls (NDW) and transverse DWs (TDW). TDW is formed at the bottom of the domain, and it corresponds to a dark contrast. In $H = 0.7$ kOe, the TDW is formed in the upper part of the domain, then it extends to the middle of NDW. The domain walls are separated by an antivortex. At fields above $H = 0.75$ kOe in most of the domains we observe by MFM only unipolar contrast, which indicates that all topological defects, such as vortices and antivortices, annihilate in high fields. 

The upper panels of Figs.~\ref{fig:4}(d) and~\ref{fig:4}(e) show the profiles of the magnetostatic fields created by N{\'e}el and transverse DWs, respectively. These profiles are determined from the MFM images in Fig.~\ref{fig:4}(a) and (b). The lower panels show the micromagnetically simulated profiles of the perpendicular component of the magnetization $M_z /M_s$ in the DWs in the top (red curve) and bottom (black curve) magnetic layers. From these simulations it can be seen that the contrast of the DWs in the MFM image is determined by the distribution of  $M_z$ components of the magnetization in the domain wall. By measuring the width of the DW profiles, $\Delta\delta$, we can estimate the change in the DW width, $w$. In the case of NDW $\Delta\delta$ is defined as the distance between the maximum and minimum values of $M_z$, whereas in the TDWs it is defined as the full width at half maximum of $M_z$. Figure~\ref{fig:4}(f) shows $\Delta\delta$ as a function of magnetic field $H$.  We have also determined the DW width $w$ as a function of $H$ from micromagnetic simulations. To achieve it, we produced DW profiles: $M_y /M_s$ for the N{\'e}el and $M_x /M_s$ for transverse DW, from which we determined $w$ as the corresponding full width at half maximum. We find that as $H$ increases $w$ is reduced. This is due to the fact that with increasing $H$, the magnetization in the domains orients along the field, see the configurations of the magnetization at different fields in Fig.~\ref{fig:3}.  Note that $\Delta\delta$ obtained from the MFM images is 565 nm, whereas $w$ determined from the simulations is 390 nm. The reason for this discrepancy is the fact that the MFM determines the profile of magnetostatic field, whereas the modeling determines the actual DW width.

\begin{figure}[ht]
\includegraphics[width=0.99\linewidth]{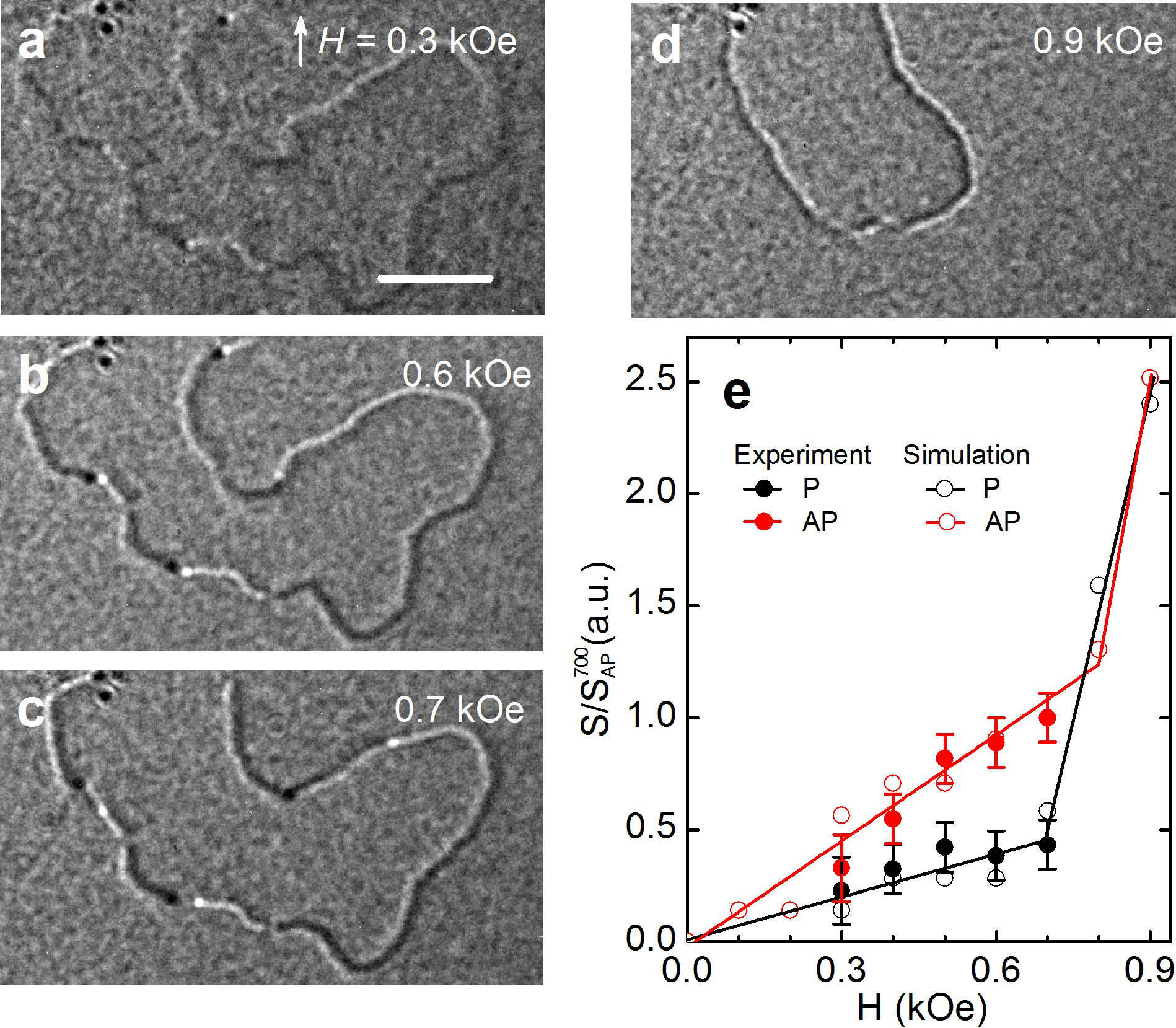}
\caption{(a) -- (d) Lorentz images of domain structure at different magnetic fields. (e) Average distance $S$ between the vortices in parallel and antiparallel configurations (both experimental data and simulations). The distance is normalized by the average experimental distance at $H=0.7$ kOe for the AP configuration.}
\label{fig:5}
\end{figure}

\textbf{Dynamics of bi-merons.} Next we study the behavior of the bi-vortices/antivortices in increasing magnetic field. Figures~\ref{fig:5}(a)--(d) show the Lorentz images of the domain structure at different fields. In the absence of magnetic field, the domain structure contrast is not present, see for details Fig.~2 of the Supplementary Material. In $H = 0.3$ kOe, the domain boundaries appear of dark and light contrast, which are separated by bi-vortices and bi-antivortices. With the increase of the field up to $H = 0.6$ kOe the contrast from vortices becomes sharp. One can see that the vortices of different chiralities branch off. At $H = 0.7$ kOe, the bi-vortices begin to move along the DW and annihilate with bi-antivortices. This process goes though the formation and consecutive annihilation of a composite bi-skyrmion formed by a pair of bi-vortex and bi-antivortex of opposite polarizations,\cite{Tretiakov07} such that in each SAF's layer their fractional topological charges $1/2$ add up to the full skyrmion charge $1$. This annihilation changes the type of DWs, their polarity and the domain shape. At $H = 1$ kOe the DWs completely disappear. These processes are clearly confirmed by our micromagnetic simulations presented in Supplementary Movies 1 and 2.
  
We have also analyzed the distance $S$ on which the bi-vortices of different chirality move as a function of magnetic field. Two groups of bi-vortices have been identified that shift differently. In both cases, the maximum displacement is achieved in field $H = 0.7$ kOe, for the first group it is $S_{\rm{AP}} = 268 \pm 30$ nm and for the second $S_{\rm{P}} = 116 \pm 30$ nm. A further increase in the field leads already to the motion of the DWs and finally their annihilation. 

The analysis of our simulations shows that the cores of the bi-vortices in the DWs move with respect to each other with increasing field. The bi-vortices (and bi-antivortices) move along the DWs with rather high velocity, so the dynamics cannot be observed in our experiments, but we have been able to reproduce them in micromagnetic simulations. Our simulation results indicate that the maximum displacement occurs for bi-vortices of opposite polarities (the antiparallel direction of $M_z$ components, $V_{\rm{AP}}$), and the minimum occurs for bi-vortices with the same polarity  ($V_{\rm{P}}$), see Fig.~\ref{fig:5}(e). This is associated with the dipolar interaction of vortex cores at separations larger than the core size, which is greater for the bi-vortices with the same polarity. \cite{Cherepov2012, Sluka2015} 

To compare the results of measurements and calculations of $S$, we have conducted a valuation of the experimental data at the maximum $S_{\rm{AP}}^{0.7} = 268 \pm 30$ nm and the corresponding simulated value $S_{\rm{AP}}^{sim} = 28.3$ nm. The results are shown in Fig.~\ref{fig:5}(e). One can see that the displacement of the bi-vortices in both cases increases linearly with the field up to $\approx$ 0.7 kOe. In the fields above $H = 0.7$ kOe, the distance between the bi-vortices increases rapidly because the magnetostatic interactions are negligible.

\section*{Conclusions} 

We have investigated both experimentally and by means of micromagnetic simulations polycrystalline Co/Ru/Co trilayers with antiferromagnetic interlayer coupling. We have demostrated that in these SAF films closed domains are bounded by complex topological spin textures. These domains possess two types of boundaries: dual N{\'e}el and transverse DWs, which are coupled by either bi-vortices or bi-antivortices as their connecting elements (Fig.~\ref{fig:DW_blocks}). By increasing magnetic field we have observed the evolution of these spin textures, which has been shown to be governed by the motion of topological defects (magnetic bi-merons) along the domain walls.  

Our finding of novel combinations of topological domain walls in synthetic AFM structures points to a new route of exploring  topologically protected and compact spin textures. We have demonstrated that these DWs may be controllably manipulated by magnetic fields and potentially driven by electric or spin currents. Additional advantage offered by these structures in SAFs is that the dipolar interactions of the two ferromagnetic layers nearly cancel each other leading to significantly reduced parasitic effect of stray fields. This study offers the perspective on new types of experiments and nanodevices, where one can move bits of information associated with the DWs of different type and polarity of vortices (antivortices) in a precise and controllable fashion in synthetic antiferromagnets.

\section*{Methods}
\subsection*{Experimental Procedure}

The  films of Co/Ru/Co were prepared by magnetron sputtering with direct current in the working gas atmosphere of Ar ($P_{\rm{Ar}} = 4\times 10^{-3}$ Torr). The base pressure in the chamber was $10^{-9}$ Torr. The films were deposited at room temperature. Prior to the deposition on the substrate of naturally oxidized silicon Si/SiO$_2$  we sputtered Ta buffer layer of  5 nm thickness. The deposition rate of Co and Ru was 0.45 and 0.3 \AA/s, respectively. In this work we explored the Co/Ru/Co film with the Co thicknesses $t_{\rm{Co}} = 10$ nm and Ru layer thickness $t_{\rm{Ru}} = 0.9$ nm, which corresponds to the first antiferromagnetic maximum. 

\subsection*{Magnetization Measurements and Imaging}

The magnetic properties of the films were investigated by the longitudinal magneto-optical Kerr effect, the magnetic structure was studied by the MFM and Lorentz microscopy. For the electron microscopy studies, we deposited the films on the standard Formvar film on Grids in a single cycle with the samples on the substrate of Si/SiO$_2$. 

\subsection*{Micromagnetic Simulations}

Micromagnetic simulations were performed using MuMax3 program. \cite{MuMax}
For the simulations we used the following parameters: $M_s = 1260$ G, $A = 3 \times 10^{-6}$ erg/cm, uniaxial anisotropy $K_u= 5 \times 10^4$ erg/cm$^3$ (the value of $K_u$ was determined experimentally for the single layer Co films for the thickness of 10 nm), and $J_{\rm{in}} = -1.4$ erg/cm$^2$. The size of the elementary cells of the Co layers was taken $4 \times 4 \times 10$ nm$^3$. The Co layers were AFM coupled via Ru layer, which was modeled by the indirect AFM coupling with the constant $J_{\rm{in}}$. We simulated film of area $4 \times 4$ $\mu$m$^2$ with periodic boundary conditions. To compare the results of our experiment with the simulations, we used calculated spin configurations to simulate images of MFM \cite{MuMax} and Lorentz microscopy.\cite{Walton2013}

\section*{ASSOCIATED CONTENT}
\subsection*{Supporting Information}
Supplementary Movie 1: Domain wall motion in a synthetic antiferromagnet with increasing magnetic field. The magnetic field is applied perpendicular to the plane ($xy$-plane).  In-plane components of the magnetization at the sites are represented by arrows in a system of $100 \times 120$ sites for the time range from $t = 0$ to $t = 3$ ns. The numerical simulations are performed in the film of area $4\times4 \mu$m$^2$ and cell size $4\times4\times10$ nm$^2$. The periodic boundary conditions are applied on the edges. The material parameters of the film were as follows: the saturation magnetization $M_s = 1260$ G, exchange stiffness $A = 3\times10^{-6}$ erg/cm, uniaxial anisotropy $K_u= 5 \times 10^4$ erg/cm$^3$, indirect exchange coupling constant $J_{\rm{in}} = -1.4$ erg/cm$^2$, Gilbert damping $\alpha = 0.02$, and magnetic field $B = 320$ Oe. Supplementary Movies 1 and 2: full simulation time is 3 ns.
Supplementary Movies 3 and 4: (zoom1) simulation time is 1 ns.
Supplementary Movies 5 and 6: (zoom2) full simulation time is 3 ns.

\section*{Acknowledgements}
We thank G.\,S.\,D. Beach for helpful discussions. This work was supported by the Russian Ministry of Education and Science (under the state tasks 3.5178.2017/8.9 and 3.4956.2017) and by Act 211 Government of the Russian Federation (contract No.\,02.A03.21.0011). O.\,A.\,T. acknowledges support by the Grants-in-Aid for Scientific Research (Grants No.\,25247056, No.\,17K05511, and No.\,17H05173) from MEXT, Japan. This project was supported by JSPS and RFBR (Grant 17-52-50060) under the Japan - Russian Research Cooperative Program. A.\,S.\,S. thanks the Brain Pool Program (172S-2-3-1928) through the Korean Federation of Science and Technology Societies (KOFST) funded by the Ministry of Science, ICT and Future Planning.

\section*{Author contributions}
A.V.O., L.A.C., A.S.S. and O.A.T. planned and supervised the study. A.G.K., A.V.O. and A.S.S. designed the experimental set-up. A.G.K., A.V.O., A.S.S., V.S.P. and E.V.P. fabricated the devices, performed the experiments, and collected data. A.G.K., A.V.O. and O.A.T. planned and performed the simulations. A.G.K., A.V.O., A.S.S. and O.A.T. analysed the data. A.V.O., A.S.S. and O.A.T. wrote the manuscript. All authors discussed the results.

\section*{Additional information}

\textbf{Competing financial interests:} The authors declare no competing financial interest.

\bibliography{micromagnetics}

\end{document}